\documentclass[prd,11pt]{revtex4-1}

\usepackage{amssymb}

\usepackage{physics}
\usepackage{bm}
\usepackage{amsmath}    % need for subequations
\usepackage{graphicx}   % need for figures
\usepackage{verbatim}   % useful for program listings
\usepackage{color}      % use if color is used in text
\usepackage{subfigure}  % use for side-by-side figures
\usepackage{hyperref}   % use forhttps://de.overleaf.com/project/5ddbf14b7a2b8700016cd45a hypertext links, including those to external documents and URLs
\definecolor{amaranth}{rgb}{0.9, 0.17, 0.31}
\definecolor{purple(munsell)}{rgb}{0.62, 0.0, 0.77}
\definecolor{americanrose}{rgb}{1.0, 0.01, 0.24}
\definecolor{palatinateblue}{rgb}{0.15, 0.23, 0.89}
\definecolor{royalblue(web)}{rgb}{0.25, 0.41, 0.88}
\definecolor{hanpurple}{rgb}{0.32, 0.09, 0.98}
\definecolor{beaublue}{rgb}{0.74, 0.83, 0.9}
\definecolor{carminered}{rgb}{1.0, 0.0, 0.22}
\definecolor{brightpink}{rgb}{1.0, 0.0, 0.5}
\definecolor{vividviolet}{rgb}{0.62, 0.0, 1.0}

\hypersetup{ linktoc=all,
    colorlinks, linkcolor={palatinateblue},
    citecolor={brightpink}, urlcolor={amaranth}}

\newcommand{\be}{\begin{equation}}
\newcommand{\ee}{\end{equation}}
\newcommand{\bs}{\begin{split}} 
\newcommand{\bea}{\begin{eqnarray}}
\newcommand{\eea}{\end{eqnarray}}

\begin{document}

\title{\Large Precession of perihelia in the Fisher metric}

\author{Nosratollah Jafari}\email{nosrat.jafari@gmail.com }

\affiliation{Department of Physics, Nazarbayev University,\\Kabanbay Batyr Ave 53, Nur-Sultan, 010000, Kazakhstan.}

\begin{abstract}

We study the precession of perihelia in the Fisher metric. Fisher metric is the solution of the Einstein's Equations with a  massless scalar field as a
coupling. We find an expression for the precession of perihelia in this metric. This expression contains general relativistic term for the precession of the perihelia and also an additional term which depends on the scalar field. Also, we obtain an upper bound on scalar charge $\sigma$ by using the observational value of the precession of perihelia for the Mercury planet and the discrepancy between this value and the general relativistic value.

 \vspace{13cm}


\end{abstract}

\maketitle

\newpage

\section{Introduction}

Adding a scalar field, minimally coupled to the gravity is a famous area of investigation.  The inflationary solutions are already some parts of the
mainstream studies cosmology.  Another natural area of investigation would be the spherical symmetric solutions, that is study the effect of a scalar field on the gravitational field of a spherical object.
An early attempt for finding solution to this problem was made by the Russian physicist, I.~Z.~Fisher in 1948 \cite{Fish}.
Later, in 1968 A.~I.~Janis, E.~T.~Newman, and J.~Winicourt studied the problem. They found a solution, known as the JNW's solution \cite{JNW}.
In 1981, M.~Wyman found a solution which is more simple than the JNW's solution \cite{Wym}.
The problem has been reconsidered by A.~G.~Agnese and M.~La~Camera in 1985 \cite{Agnes}. They have some discussions on the nature of singularity in
 this solution. This singularity was further investigated in papers by P.~S.~Joshi  and his collaborators \cite{Josh1, Josh2}.
In 1997, K.~S.~Virbhadra showed that the Wyman's solution and JNW's
solution are the same \cite{Vib1}.

 In this paper, we will consider the solution which has been introduced by Agnese La~Camera.  This solution is more similar to the
 Schwarzschild's solution, and it is easy to deal with than JNW's solution.
We will call this solution as the Fisher solution.

Observational effects of the Fisher solution are interesting. For example,  K.~S.~Virbhadra, D.~Narasimha, and S.~M.~Chitre have studied
 the deflection of light by the Fisher metric \cite{Vib2}. Also, Virbhadra and C.~R.~Keeton have studied the effect of time delay in this 
 metric \cite{Vib3}.

Recently, J.~B.~Formiga has studied the observational effect of a massless scalar field in the solar system experiments \cite{Form1}.
By experiments Formiga has meant observations of the sending signals to the Sun or the spectral shift of the light. He has argued that observational effects
 of the massless scalar field can not be detected in the solar system experiments. Formiga and his collaborators has formulated a general scalar
 tensor theory action for gravity \cite{Form2}. In this approach, they have regarded the fisher solution as a special case of Weyl's geometry and they
 have discussed that this metric is invariant under Weyl transformations and  therefore is invisible in the observations.

Fisher metric is the solution of the Einsetn's equation for the spherically symmetric metric with a scalar field coupling. We should solve a D'Alembert's equation for the scalar field. The solution of this equation for the scalar field has a logarithmic dependency. In the large distances scalar field has an inverse of radial distance dependency and in analogy with the Coulomb's field, we can introduce the scalar charge.

In this paper, we study the precession of perihelia in the Fisher metric and we obtain an expression for the precession of perihelia. This expression contains general relativistic  term for the precession and also an additional term which depends on the scalar field. From which we obtain an upper bound on value of scalar charge. It is very hard to observe the effect of scalar field in the solar system. Because of this matter some screening mechanisms have been developed to explain invisibility  \cite{Khou1, Khou2, Khou3, Mot}.

\section{Equations of motion in the Fisher metric}

We are looking for spherically symmetric metric which has a massless scalar field as a coupling.  This metric is known as the Fisher solution \cite{Fish}.
 Here, we consider the solution which has been found  later by Agnese and La~Camera \cite{Agnes}. It reads
\be  \label{eq1} ds^2=  \Big(1- \frac{2\eta}{r}  \Big)^{M/\eta} dt^2 - \Big(1- \frac{2\eta}{r}  \Big)^{-M/\eta} dr^2  -
              \Big(1- \frac{2\eta}{r}  \Big)^{1-M/\eta} r^2 ( d\theta^2 + \sin^2\theta d\phi^2 ), \ee
where
\be \label{2} \eta= \sqrt{ M^2 + \sigma^2 }, \ee
$M$ being the mass and $\sigma$  the scalar charge \cite{Agnes}; and we have used units such that $G = 1$ and $c = 1$.

The equation of motion for the massless scalar field is
 \be  \square^2 \phi =0.   \ee
A static solution for this equation which has spherical symmetry is
\be  \label{4}  \phi(r)= \frac{\sigma}{2 \eta}\ln \Big(1- \frac{2\eta}{r} \Big).    \ee
Using this solution, we can rewrite the Fisher metric Eq.~(\ref{eq1}) as
\be  \label{5}  ds^2= \Big( 1- \frac{2\eta}{r}  \Big) e^{-2\tau\phi} dt^2 -
                     \Big( 1- \frac{2\eta}{r}  \Big)^{-1}e^{2\tau\phi}  dr^2  -r^2 e^{2\tau\phi}  d\Omega^2,   \ee
 in which
   \be        \tau= \sqrt{\frac{\eta-M}{\eta + M}}\simeq \frac{\sigma}{2\,M}+ O(\sigma^2). \ee
   This form of the Fisher metric is a convenient form for decoupling the effect of the scalar field $\phi$ and the the modified mass  $\eta$.
   Also, the  $\tau$ parameter is a measure for showing the deviation of the Fisher metric from the Schwarzschild metric---the bigger $\tau$ shows bigger
   deviation. In fact, the bigger value of $\tau$ shows that the effect of the scalar field can not be negligible.

To study motion of a particle in metric Eq.~(\ref{5}) we consider a general spherically symmetric form  for the metric
   \be  \label{7}  ds^2= B(r) dt^2 - A(r) dr^2 - D(r) r^2 d\theta^2 - D(r) r^2 \sin^2 \theta d\psi^2 ,            \ee
   where $A(r)$,  $B(r)$, and $D(r)$ are functions of $r$.
   The equations of motion for a test particle are as follows.
   \be    \label{8}  B(r) \frac{dt}{d \tau}= k,  \ee
   \be                      \ddot{r} + \frac{B'}{2 A}  \dot{t}^2  +  \frac{A'}{2 A}  \dot{r}^2  - \frac{D'}{2 A} r^2 \dot{\psi}^2 -
   \frac{D}{ A} r \dot{\psi}^2  =0 , \ee
   \be  \label{9}     D(r) r^2 \frac{d\psi}{d \tau}=J, \ee
where $k$ and $J$ are constants;  $\tau$  is the proper time;  a dot denote the derivative with respect to this time; and
a prime denote the derivative with respect to $r$.  Using these equations, it is straightforward to write them as
 \be  \label{16}  \frac{A}{D^2 r^4} \Big(  \frac{dr}{d\psi} \Big)^2 +  \frac{1}{ D r^2 }  -   \frac{1}{ J^2 B} = -\frac{E}{J^2},   \ee
   \be        r^2 \frac{d\psi}{dt}=\frac{J B }{D}, \ee
   \be           ds^2= E B^2 dt^2,     \ee
   where $E$ is the ``energy'' of the particle.
   The solution for (\ref{16}) will be determined by integration as
      \be \label{19}   \psi=   \pm \int  \frac{A^{1/2}(r) dr }{D r^2  \Big[   1/(J^2 B) - E/J^2- 1/(D r^2)    \Big]^{1/2}  }    \ee

 \section{Precession of Perihelia}

In the remaining of this paper, we write index ``F"  for Fisher and index ``S"  for the Schwarzschild.
Thus, we have

\be    A_F(r)= \Big(1- \frac{2 \eta}{r}  \Big)^{-1} e^{2 \tau \phi}, \qquad B_F(r)= \Big(1- \frac{2 \eta}{r}  \Big) e^{-2 \tau \phi},
 \qquad D_F(r)= e^{2 \tau \phi}, \ee
 and
 \be    A_S(r)= \Big(1- \frac{2 M}{r}  \Big)^{-1} , \qquad B_S(r)= \Big(1- \frac{2 M}{r}  \Big), \qquad D_S(r)=1. \ee

 By expanding $A_F(r)$,  $B_F(r)$, and $ D_F(r) $ in powers of $1/r$ we rewrite the Fisher metric as
 \be  \label{17}  ds^2=  \Big(     1- \frac{2M}{r}   +  \frac{\sigma^2}{r^2}  \Big) dt^2 -  \Big(     1- \frac{2M}{r}   +
 \frac{\sigma^2}{r^2}  \Big)^{-1}   dr^2  -  \Big(     1- \frac{ \sigma^2}{Mr}   -  \frac{\sigma^2}{r^2}  \Big) r^2 d \Omega^2.  \ee
This form is similar to  the Reissner-Nordstrom metric, except for the sign of the second order terms.

At perihelia and aphelia, $r$ reaches its minimum $r_{-}$ and maximum $r_{+}$  and at both points $  dr/d\psi$ vanishes, so Eq.~(\ref{16} ) gives
 \be     \frac{1}{ D_F(r_{\pm})  r_{\pm}^2  }   - \frac{1}{ J_F^2 B_F(r_{\pm} ) } =  - \frac{E_F}{J_F^2}.         \ee
 From these equations we can find for the constants of motion  $E$ and $J$, the following expressions
 \be       E_F=\frac{     \frac{D_{F+} r_{+}^2  }{ B_{F+} }   -    \frac{D_{F-} r_{-}^2  }{ B_{F-} }    }{  D_{F+}  r_{+}^2 -  D_{F-}  r_{-}^2  },  \ee
 and
 \be   J_F^2 =  \frac{  \frac{1}{B_{F+}}- \frac{1}{B_{F-}}  }{ \frac{1}{ D_{F+} r_{+}^2  }   -  \frac{1}{ D_{F-} r_{-}^2  } }.     \ee
In the leading order, we have
\be \label{21}  E_F= E_S \Big[  1 +    \frac{2\sigma^2}{( r_- + r_+ )^2}   \Big]    \ee
\be \label{22}   J_F^2=  J_S^2  + \sigma^2    \ee
 The angle swept out from $ r_{-} $ by the position vector as $r$ increases from $r_{-} $ can be find by using the integral Eq.(\ref{19}) as
  \be   \psi(r)= \psi (r_{-})   +  \int_{r_{-}}^{r}  \frac{A_F^{1/2}(r)  }{ D_F(r)  }
  \Big[ \frac{1}{ J_F^2 B_F(r)  } -  \frac{E_F}{ J_F^2  }  - \frac{1}{D_F(r)  r^2   }     \Big]^{-1/2}    \frac{dr}{ r^2  } .  \ee
By using (\ref{21}) and (\ref{22}) we can write
\begin{eqnarray}
\Big[ \frac{1}{ J_F^2 B_F(r)  } -  \frac{E_F}{ J_F^2  }  - \frac{1}{C_F(r)  r^2   }
\Big] \simeq  \Big[ \frac{1}{ J_S^2 B_S(r)  } -  \frac{E_S}{ J_S^2  }  - \frac{1}{  r^2   }     \Big] \nonumber\\
  + \sigma^2 \Big( \frac{1}{J_S^2 B_S } \Big[   \frac{1}{ r^2 } -    \frac{1}{ J_S^2 } \Big] -\frac{E_S}{J_S^2 }
  \Big[  2 \Big(    \frac{r_+ -  r_-}{r_+^2 - r_-^2 }   \Big)^2 -\frac{1}{ J_S^2}   \Big]  \Big),
    \end{eqnarray}
which  in the leading order reads
   \be      \Big[ \frac{1}{ J_F^2 B_F(r)  } -  \frac{E_F}{ J_F^2  }  - \frac{1}{C_F(r)  r^2   }     \Big] \simeq  \Big[ \frac{1}{ J_S^2 B_S(r)  } -
    \frac{E_S}{ J_S^2  }  - \frac{1}{  r^2   }     \Big]   +     O \Big( \frac{\sigma^2}{r^2} \Big).   \ee
   By using these approximations, we have
    \be \psi(r) - \psi(r_-)  = \int_{r_-}^{r}  \Big( 1 + \frac{M}{r}  +  \frac{\sigma^2}{ Mr}   \Big) \Big[ \frac{1}{ J_S^2 B_S(r)  } -
    \frac{E_S}{ J_S^2  }  - \frac{1}{  r^2}     \Big]^{-1/2} \frac{ dr}{ r^2 }  +   O \Big( \frac{\sigma^2}{ r^2 } \Big).   \ee
    Thus, precession per revolution will be
    \be \label{27}  \Delta \psi = \frac{6\pi M }{ L  }   +    \frac{2 \pi \sigma^2  }{ M L  }.   \ee
    Here $L$ is the semi-latus rectum for the planetary orbit. The expression $ 6 \pi M/ L  $ is the general relativistic value for the precession
     of perihelion and we will denote it with $ \Delta \psi_{GR}$. Correction to the general relativistic value is of 
     $ (\sigma/M)^2 $ order as mentioned previously without proof in \cite{Agnes, Form1}.

 For Mercury the value of $ \Delta \psi_{GR}$ is about $ 42.94 $  arcsec per century, and $L \simeq 5.56 \times 10 ^{10} $ m. The observed value of the precession of perihelion of the planet Mercury is  $ \Delta \psi_{Obs} = 43.11 \pm 0.21 $  arcsec per century \cite{Shap}. Thus, the
  difference $ \Delta \psi_{Obs} - \Delta \psi_{GR} = 0.17 $   arcsec per century can be attributed to other
   effects such as the effect of the scalar filed. Form which, we will find \be  \sigma \simeq 1.6 \times 10 ^2  \textrm{m},  \ee as an upper bound on the scalar charge.

    This expression for the precession of perihelia (\ref{27})  is very similar to the expression for the precession of perihelia in the Reissner-Nordstrom
    metric \cite{Boeh}. This similarity is an expectable result , because the Fisher metric in the second order of the approximation in $\sigma$  as
    in (\ref{17}) is very similar to the Reissner-Nordstrom metric.

\vspace{3cm}


\newpage


\begin{thebibliography}{99}


\bibitem{Fish} Fisher,  Zh. Eksp. Teor. Fiz. 18, 636 (1948). Translation from Russian: " Scalar mesostatic field with regard for gravitational effects", gr-qc/9911008.

\bibitem{JNW} A. I. Janis, E. T. Newman and J. Winicourt, Phys .Rev. Lett. \textbf{20}, 878 (1968).

\bibitem{Wym} M. Wyman, Phys. Rev.\textbf{ D24}, 839 (1981).

\bibitem{Vib1} K.S. Virbhadra, Int. J. Mod. Phys.\textbf{ A12}, 4831 (1997).

\bibitem{Agnes} A. G. Agnese, Phys. Rev. \textbf{D31}, 1280 (1985).

\bibitem{Josh1}  P. S. Joshi, D. Malafarina, Int. J. Mod. Phys. \textbf{D20} 2641 (2011)

\bibitem{Josh2}   S. Bhattacharya, P. S. Joshi, Mod. Phys. Lett. \textbf{A26}, 1281 (2011)


\bibitem{Vib2} K. S. Virbhadra, D. Narasimha, and S. M. Chitre,  Astron.Astrophys. \textbf{337}, 1 (1998).

\bibitem{Vib3}  K.S. Virbhadra, C.R. Keeton, Phys.Rev. \textbf{D77}, 124014 (2008).


\bibitem{Form1} J. B. Formiga, Phys. Rev. \textbf{D83}, 087502 (2011).

\bibitem{Form2}   T. S. Almeida, M. L. Pucheu, C. Romero and J. B. Formiga, Phys. Rev. \textbf{D89}, 064047 (2014).

\bibitem{Wein} S. Weinberg,Gravitation and Cosmology, John Wiley
Sons, Inc. (1972).

\bibitem{Will}, C. M. Will, Liv. Rev. Rel. 17, 4 (2014); arxiv: 1403.7377.


\bibitem{Khou1} J. Khoury and A. Weltman, Phys. Rev. Lett. \textbf{93}, 171104 (2004).


\bibitem{Khou2} S.S. Gubser and J. Khoury, Phys. Rev. \textbf{D70}, 104001 (2004).

\bibitem{Khou3}   K. Hinterbichler, J. Khoury, Phys. Rev. Lett. \textbf{104}, 231301 (2010).

\bibitem{Mot}  	 D. F. Mota, D. J. Shaw,   Phys. Rev. Lett. \textbf{97}, 151102 (2006).


\bibitem{Shap} Shapiro II, Counselman C. C. and King R. W., Phys. Rev. Lett. \textbf{36}, 555 (1976).

\bibitem{Boeh}   C. G. Boehmer, T. Harko, F. S. N. Lobo, Class. Quant. Grav. \textbf{25}, 045015 (2008).



\end{thebibliography}
\end{document}